\documentclass[a4paper,fleqn]{cas-sc}

\usepackage[authoryear,longnamesfirst]{natbib}

\def\tsc#1{\csdef{#1}{\textsc{\lowercase{#1}}\xspace}}
\tsc{WGM}
\tsc{QE}

\begin{document}
\let\WriteBookmarks\relax
\def\floatpagepagefraction{1}
\def\textpagefraction{.001}

\shorttitle{Theory of viscoelasticity and sound speed}    

\shortauthors{Wang}  

\title [mode = title]{Theory of viscoelasticity and sound speed}  

\author{Fei Wang}[orcid=0000-0003-3318-3264]

\affiliation{organization={Institute of Nanotechnology (INT),
Karlsruhe Institute of Technology (KIT)},
            addressline={Hermann-von-Helmholtz-Platz 1}, 
            city={Eggenstein-Leopoldshafen},
            postcode={76344}, 
           state={Baden Wuerttemberg},
            country={Germany}}

\cortext[1]{Email: fei.wang@kit.edu}

\begin{abstract}
This work develops an energy-based framework for describing coupled viscous and elastic dynamics in continuum materials. By incorporating kinetic energy and elastic deformation into a unified free-energy formulation, coupled evolution equations for velocity, strain, and momentum are derived. The resulting model accounts for viscous dissipation through velocity diffusion and phenomenological strain relaxation, while elastic deformation contributes to momentum evolution through volumetric and shear responses. Furthermore, under homogeneous, isotropic, linear-elastic, and isentropic assumptions, the framework recovers the classical sound speed in liquids and the longitudinal and transverse wave speeds in solids. These results provide a unified energy-based perspective on viscous dissipation, elastic deformation, and wave propagation in continuum materials.
\end{abstract}



\maketitle

\section{Introduction}

In conventional continuum mechanics~\cite{Newton1687,landau1987fluid}, the motion of a material is described by the balance of linear momentum. For a fluid particle subjected only to pressure forces, this equation takes the form
\begin{equation}
\rho\frac{\mathrm D\mathbf v}{\mathrm Dt}=-\nabla p,
\end{equation}
where \(\rho\), \(\mathbf v\), and \(p\) denote the mass density, velocity, and pressure, respectively. This equation can be equivalently expressed in terms of the momentum evolution as
\begin{equation}
\frac{\mathrm D(\rho\mathbf v)}{\mathrm Dt}=-\nabla p,
\end{equation}
provided that the material density satisfies
\begin{equation}
\frac{\mathrm D\rho}{\mathrm Dt}=0.
\end{equation}
The latter condition expresses the conservation of density along material trajectories. Within the conventional formulation, the momentum balance plays a central role in describing and determining the motion of the continuum.

In previous work~\cite{wang2026density}, we proposed an alternative energy-based interpretation of continuum dynamics by introducing a Lagrange multiplier $p$ associated with mass into the kinetic-energy formulation,
\begin{equation}
E=\rho\mathbf v^2+p.
\end{equation}
By decomposing the differential of the system energy into dissipative and nondissipative contributions,
\begin{equation}
\mathrm dE=
\underbrace{\rho\mathbf v\cdot\mathrm d\mathbf v}_{\text{dissipative}}
+
\underbrace{\mathbf v\cdot\mathrm d(\rho\mathbf v)+\mathrm dp}_{\text{nondissipative}},
\end{equation}
we derived a velocity-diffusion equation from the dissipative contribution and a kinetic Gibbs--Duhem relation from the nondissipative contribution. Within the proposed framework, the latter is equivalent to the momentum evolution equation. This formulation assigns distinct roles to velocity and momentum evolution: the velocity equation governs the dissipative dynamics and thereby determines the motion of the material, whereas the momentum relation determines the Lagrange multiplier \(p\), which is identified with the hydrodynamic pressure. This energy-based framework has subsequently been extended to fluid--electromagnetic coupling~\cite{wang2026thermodynamics} and quantum and relativistic dynamics~\cite{wang2026relativistic}, demonstrating the applicability of the kinetic Gibbs--Duhem relation to a broader range of physical systems.

The present work extends this framework to solid phases by incorporating elastic deformation and viscous dissipation into a unified energy-based formulation. Unlike fluids, solid materials can sustain shear stresses and store elastic energy through deformation. Their dynamical behavior therefore involves not only the evolution of the velocity field but also that of the strain field. In classical solid mechanics, the constitutive description of such materials is commonly formulated in terms of free-energy functions and dissipation mechanisms. Variational approaches provide a systematic framework for describing the coupled thermomechanical behavior of dissipative solids~\cite{Yang2006}. More recent developments have further explored thermodynamic potentials and internal variables in linear viscoelastic composites~\cite{Lahellec2024}, as well as energy-based approaches to the characterization of linear viscoelastic materials~\cite{Bonnet2024}. These studies highlight the importance of consistently describing elastic energy storage and viscous dissipation within continuum constitutive theories.

To this end, we construct a free-energy functional comprising kinetic and elastic contributions, together with a Lagrange multiplier associated with the relevant conservation constraints. By decomposing the energy evolution into dissipative and nondissipative contributions, we derive coupled evolution equations for velocity, strain, and momentum. Specifically, the dissipative contributions give rise to a velocity-diffusion equation governed by viscosity and density, together with a phenomenological strain-diffusion equation describing the relaxation of shear and volumetric deformation. The nondissipative contribution determines the momentum evolution through the volumetric and shear components of the elastic response. Existing viscoelastic formulations have addressed compressible and incompressible finite deformation through constitutive and internal-variable approaches~\cite{Wijaya2023,Liu2025}. The present approach differs in its use of the kinetic Gibbs--Duhem relation to distinguish the dissipative evolution of velocity and strain from the nondissipative momentum evolution. Taken together, these equations provide an energy-based description of the coupling between viscous dissipation and elastic deformation in continuum solids.

Within this framework, we further investigate longitudinal and transverse wave propagation in solids, as well as sound propagation in liquids. Under homogeneous, isotropic, linear-elastic, and isentropic assumptions, the formulation recovers the classical wave speeds in terms of the bulk modulus, shear modulus, and mass density. These results establish a connection between the proposed energy-based framework and classical wave propagation in both fluid and solid phases.

The remainder of this paper is organized as follows. Section~2 introduces the definitions and assumptions used throughout the analysis. Section~3 develops the free-energy formulation and derives the coupled evolution equations for velocity, strain, and momentum. Section~4 investigates sound ve propagation in fluid and solid phases. Section~5 examines the transformation properties of the momentum evolution equation and its behavior under Galilean transformations. Finally, Section~6 summarizes the main findings and conclusions.

\section{Definition}
To describe the material motion, we adopt a Lagrangian framework, in which the position $\mathbf r$ refers to a material point and evolves with time. The velocity of the material point is therefore defined as
\begin{equation}
 \mathbf v
 =
 \frac{\mathrm D\mathbf r}{\mathrm Dt}.
\end{equation}
Here, the time derivative is taken while following the same material point. Consequently, the Lagrangian description naturally connects the evolution of the material position, velocity, density, and stress. 

The Galilei transformation is defined as 
\begin{align}
 x^\prime&=x+ Ut=x+x_0,\\
 t^\prime&=t+t_0,
\end{align}
where $x_0$ and $t_0$ are constants, $U$ is an arbitrary constant velocity.

In the Lagrangian framework with velocity much less than that of light, we define the incompressible condition as 
\begin{equation}
 \frac{\mathrm D (\delta V)}{\mathrm D t}=0,
\end{equation}
where $\delta V $ is an infinitesimal material volume element.
This definition differs from another one by using the density as the incompressible variable~\cite{Lundgren2024}.
Since the rate of change of a material volume element is given by
\begin{equation}
\frac{\mathrm D(\delta V)}{\mathrm Dt}
=(\nabla\cdot\mathbf v)\delta V,
\end{equation}
the incompressibility condition directly yields
\begin{equation}
\nabla\cdot\mathbf v=0.
\end{equation}

\section{Theory}

\subsection{Free-energy formulation and evolution equations}

We consider a homogeneous, isotropic, linearly elastic solid-fluid subjected to external loading. Within the small-strain framework, the elastic free energy is expressed in terms of the strain tensor $\boldsymbol{\varepsilon}$ and the fourth-order elasticity tensor $\boldsymbol{C}$ as~\cite{landau1982mechanics,Ulloa2022}
\begin{equation}
 F_0=\int_V
 \boldsymbol{\varepsilon}:\boldsymbol{C}:
 \boldsymbol{\varepsilon} \delta V.
\end{equation}

For an isotropic material, assuming constant elastic moduli, the elastic free energy can be simplified to~\cite{Anand2022}
\begin{equation}
\begin{aligned}
 F_0
 &=\int_V
 \boldsymbol{\varepsilon}:\boldsymbol{C}:
 \boldsymbol{\varepsilon}\delta V=\int_V\left[
 \lambda(\operatorname{tr}\boldsymbol{\varepsilon})^2
 +2\mu\boldsymbol{\varepsilon}:\boldsymbol{\varepsilon}
 \right]\delta V,
\end{aligned}
\end{equation}
where $\lambda$ and $\mu$ are the Lamé parameters, representing the elastic response to volumetric and shear deformation, respectively.

By incorporating the kinetic free energy, the total system energy is extended to include both elastic deformation and material motion:
\begin{equation}
 F_0=\int_V\left[
 \lambda(\operatorname{tr}\boldsymbol{\varepsilon})^2
 +2\mu\boldsymbol{\varepsilon}:\boldsymbol{\varepsilon}
 +\rho\mathbf v\cdot\mathbf v
 \right]\delta V.
\end{equation}

The elastic contribution can also be interpreted in terms of the energy associated with microscopic excitations, such as phonons. In this interpretation, the microscopic kinetic energy density is related to the phonon energy through
\begin{equation}
 \hbar\omega=m_p\mathbf v_p^2=\int_V\left[
 \lambda(\operatorname{tr}\boldsymbol{\varepsilon})^2
 +2\mu\boldsymbol{\varepsilon}:\boldsymbol{\varepsilon}
  \right]\delta V,
\end{equation}
where $m_p$ and $\mathbf v_p$ denote the phonon mass and characteristic velocity, respectively. This relation is understood as an effective energy-density representation, with the appropriate normalization of the phonon energy.

To ensure the Galilean invariance of the system energy (see Section 5), we introduce a Lagrange multiplier $p$ associated with the mass density $\rho$. The resulting constrained free-energy functional is written as
\begin{equation}
\begin{aligned}
 F
 &=\int_V\left[
 \rho_{p}\mathbf v_p^2
 +\rho\mathbf v\cdot\mathbf v+p
 \right]\delta V\\
 &=\int_V\left[
 \lambda(\operatorname{tr}\boldsymbol{\varepsilon})^2
 +2\mu\boldsymbol{\varepsilon}:\boldsymbol{\varepsilon}
 +\rho\mathbf v\cdot\mathbf v+p
 \right]\delta V.
\end{aligned}
\end{equation}

In the non-relativistic regime, where $|\mathbf v|\ll c$ ($c$: speed of light in vacuum), the volume element is invariant under Galilean transformations. Assuming a fixed integration domain, $\delta V$ is therefore unchanged during the variation considered here, i.e., $\mathrm d (\delta V)=0$. The total differential of the system energy can consequently be expressed as
\begin{equation}
\begin{aligned}
 \mathrm dF=\int_V\Big\{&
 \lambda(\operatorname{tr}\boldsymbol{\varepsilon})
 \mathrm d(\operatorname{tr}\boldsymbol{\varepsilon})
 +(\operatorname{tr}\boldsymbol{\varepsilon})
 \mathrm d[\lambda(\operatorname{tr}\boldsymbol{\varepsilon})]
 +2\mu\boldsymbol{\varepsilon}:
 \mathrm d\boldsymbol{\varepsilon}
 +\boldsymbol{\varepsilon}:
 \mathrm d(2\mu\boldsymbol{\varepsilon})
 +\rho\mathbf v\cdot\mathrm d\mathbf v
 +\mathbf v\cdot\mathrm d(\rho\mathbf v)
 +\mathrm dp
 \Big\}\delta V.
\end{aligned}
\end{equation}

We decompose the energy variation into dissipative and nondissipative contributions. The dissipative terms describe the relaxation and diffusion of the velocity and strain fields, whereas the nondissipative terms determine the corresponding momentum balance and constitutive constraints.

\subsection{Velocity diffusion and the kinetic Gibbs–Duhem relation}

We first consider the kinetic contribution to the energy variation:
\begin{equation}
 \rho\mathbf v\cdot\mathrm d\mathbf v
 +\mathbf v\cdot\mathrm d(\rho\mathbf v) +\mathrm dp.
\end{equation}
The first term, $\rho\mathbf v\cdot\mathrm d\mathbf v$, is associated with the dissipative evolution of the velocity field. By applying the dissipation principle~\cite{zhang2024multi}, we obtain a diffusion-type evolution equation:
\begin{equation}
 \frac{\mathrm D\mathbf v}{\mathrm Dt}
 =
 \nabla\cdot D_v\nabla\boldsymbol{\mu}_v.
 \label{eq:14-5}
\end{equation}
Here, $D_v$ denotes the velocity mobility, defined as
\begin{equation}
 D_v=\frac{\eta}{\rho^2},
\end{equation}
where $\eta$ is the dynamic viscosity. The corresponding chemical-velocity potential is defined by
\begin{equation}
 \boldsymbol{\mu}_v=\rho\mathbf v.
\end{equation}
Thus, the velocity evolution is governed by the spatial diffusion of the momentum-density field, with the mobility determined by the material viscosity and density.

The remaining nondissipative contributions lead to a kinetic Gibbs–Duhem relation~\cite{wang2026density}:
\begin{equation}
 \mathbf v\cdot\mathrm d(\rho\mathbf v)+\mathrm dp=0.
\end{equation}
This relation connects the variations of the kinetic momentum density and the Lagrange multiplier $p$. Its spatial counterpart gives the momentum evolution equation~\cite{wang2026density}
\begin{equation}
 \frac{\mathrm D(\rho\mathbf v)}{\mathrm Dt}
 =-\nabla p.
\end{equation}
This equation describes the momentum balance associated with the kinetic contribution, in which the gradient of the Lagrange multiplier acts as the corresponding driving force.
The physical meaning of the Lagrange multiplier $p$ is the hydrodynamic pressure,
as discussed in Refs.~\cite{blokhuis1995young,johnsonConflictsGibbsianThermodynamics1959,gibbs1906scientific}.

\subsection{Elastic contribution to the momentum evolution}

We now consider the complete variation of the elastic and kinetic contributions to the system energy. The corresponding generalized Gibbs–Duhem relation becomes
\begin{equation}
\begin{aligned}
 \mathrm dp
 +\mathbf v\cdot\mathrm d(\rho\mathbf v)
 +\boldsymbol{\varepsilon}:
 \mathrm d(2\mu\boldsymbol{\varepsilon})
+(\operatorname{tr}\boldsymbol{\varepsilon})
 \mathrm d[\lambda(\operatorname{tr}\boldsymbol{\varepsilon})]
 =0.
\end{aligned}
\label{eq:22-9}
\end{equation}

The additional terms represent the contribution of elastic deformation to the momentum balance. In particular, the shear and volumetric components of the strain energy generate additional forces associated with spatial variations of the strain field.
Dividing by $\mathrm d x_i$ on both sides of Eq.~\eqref{eq:22-9}, 
the corresponding momentum evolution equation is obtained as
\begin{align}
 \frac{\mathrm D(\rho\mathbf v)}{\mathrm Dt}
 ={}&-\nabla p
 -\boldsymbol{\varepsilon}:
 \nabla(2\mu\boldsymbol{\varepsilon})-(\operatorname{tr}\boldsymbol{\varepsilon})
 \nabla[\lambda(\operatorname{tr}\boldsymbol{\varepsilon})]
 \notag\\
 ={}&-\nabla\cdot\left[
 2\mu\boldsymbol{\varepsilon}^2
 +\frac{\lambda}{2}
 (\operatorname{tr}\boldsymbol{\varepsilon})^2\mathbf I
 +p\mathbf I
 \right]+2\mu(\nabla\cdot\boldsymbol{\varepsilon})
 \cdot\boldsymbol{\varepsilon}.
 \label{eq:12-9}
\end{align}
This expression shows explicitly how the elastic deformation modifies the momentum evolution. The resulting force consists of the gradient of the hydrodynamic pressure, the divergence of the elastic stress-like contribution, and an additional term associated with spatial variations of the strain tensor.
Eq.~\eqref{eq:22-9} also provides a physical origin 
of the Cauchy stress tensor:
\begin{equation}
 \boldsymbol \sigma= 2\mu\boldsymbol{\varepsilon}^2
 +\frac{\lambda}{2}
 (\operatorname{tr}\boldsymbol{\varepsilon})^2\mathbf I
 +p\mathbf I.
\end{equation}

\subsection{Strain diffusion and dissipative evolution}

Finally, we consider the dissipative contributions associated with the strain energy:
\begin{equation}
 \lambda(\operatorname{tr}\boldsymbol{\varepsilon})
 \mathrm d(\operatorname{tr}\boldsymbol{\varepsilon})
 +2\mu\boldsymbol{\varepsilon}:
 \mathrm d\boldsymbol{\varepsilon}.
\end{equation}

Applying the dissipation principle~\cite{zhang2024multi,wang2023thermodynamically} to these terms leads to a diffusion-type evolution equation for the strain tensor:
\begin{equation}
\boxed{
 \frac{\mathrm D\boldsymbol{\varepsilon}}{\mathrm Dt}
 =
 \frac{\partial\boldsymbol{\varepsilon}}{\partial t}
 +\mathbf v\cdot\nabla\boldsymbol{\varepsilon}
 =
 \nabla\cdot(\mu\nabla\boldsymbol{\varepsilon})
 +\nabla(\lambda\nabla\cdot\boldsymbol{\varepsilon}).
}
\end{equation}

The first term on the right-hand side describes the diffusion of shear deformation, while the second accounts for the diffusion associated with volumetric deformation. Accordingly, the strain evolution equation incorporates both shear and volumetric relaxation mechanisms.
Together, the velocity and strain evolution equations provide a coupled description of dissipative motion and elastic deformation, thereby establishing the viscous–elastic framework developed in this work.

\section{Sound speed} 

We next validate the viscoelastic theory by reproducing sound speeds in solid and fluids. 
Combining the velocity-diffusion equation, Eq.~\eqref{eq:14-5}, with the momentum-evolution equation, Eq.~\eqref{eq:12-9}, and subtracting the latter from the former after multiplying Eq.~\eqref{eq:14-5} by \(\rho\), we obtain the following density-evolution relation:
\begin{align}
 \mathbf v\frac{\mathrm D\rho}{\mathrm Dt}=
 -\nabla p
 -\nabla\cdot\left[
 2\mu\boldsymbol{\varepsilon}^2
 +\frac{\lambda}{2}
 (\operatorname{tr}\boldsymbol{\varepsilon})^2\mathbf I
 \right]
 +2\mu(\nabla\cdot\boldsymbol{\varepsilon})
 \cdot\boldsymbol{\varepsilon}
 -\rho\nabla\cdot
 \left[\frac{\eta}{\rho^2}
 \nabla(\rho\mathbf v)\right].
\end{align}
Note that the hydrodynamic pressure $p$   is distinguished from the elastic contributions associated with the strain field.

For the analysis of elastic-wave propagation under isentropic conditions, we neglect the velocity-dissipative contribution. The density-evolution equation consequently reduces to
\begin{equation}
 \mathbf v\frac{\mathrm D\rho}{\mathrm Dt}
 =
 -\nabla p
 -\nabla\cdot\left[
 2\mu\boldsymbol{\varepsilon}^2
 +\frac{\lambda}{2}
 (\operatorname{tr}\boldsymbol{\varepsilon})^2\mathbf I
 \right]
 +2\mu(\nabla\cdot\boldsymbol{\varepsilon})
 \cdot\boldsymbol{\varepsilon}.
\label{eq:31-00}
\end{equation}

\subsection{Sound propagation in fluids}

In fluid phases, the strain-dependent elastic contributions vanish, and the density-evolution equation becomes
\begin{equation}
 \mathbf v\frac{\mathrm D\rho}{\mathrm Dt}
 =-\nabla p.
\end{equation}
Considering a one-dimensional disturbance propagating along the direction of the fluid velocity, and assuming steady motion, we have
\[
\frac{\mathrm D\rho}{\mathrm Dt}
=v\frac{\mathrm d\rho}{\mathrm dx},
\qquad
\frac{\mathrm dp}{\mathrm dx}
=\frac{\mathrm dp}{\mathrm d\rho}
\frac{\mathrm d\rho}{\mathrm dx}.
\]
It follows that
\begin{equation}
 v^2\,\mathrm d\rho=-\mathrm d p.
\end{equation}
Consequently, the sound speed is given by
\begin{equation}
\boxed{
v=\sqrt{\frac{\mathrm{d}(-p)}{\mathrm{d}\rho}}
=\sqrt{\frac{K}{\rho}},}
\end{equation}
where \(K\) is the bulk modulus, defined for a compressible medium by
\begin{equation}
\frac{\mathrm{d}(-p)}{\mathrm{d}\rho}=\frac{K}{\rho}.
\end{equation}
The negative sign in these expressions arises from the choice of \(+p\) for the Lagrange multiplier. Alternatively, the multiplier can be defined as \(-p\), in which case the negative sign is eliminated. These two sign conventions correspond to choosing the inward or outward normal direction, respectively, for the boundary of the particle under consideration.

\subsection{Longitudinal and transverse sound propagation in solids}

In solids, the density evolution is governed by the elastic contributions: 
\begin{equation}
 \mathbf v\frac{\mathrm D\rho}{\mathrm Dt}
 = -\nabla\cdot\left[
 2\mu\boldsymbol{\varepsilon}^2
 +\frac{\lambda}{2}
 (\operatorname{tr}\boldsymbol{\varepsilon})^2\mathbf I
 \right]
 +2\mu(\nabla\cdot\boldsymbol{\varepsilon})
 \cdot\boldsymbol{\varepsilon}.
 \label{eq:31-000}
\end{equation}
Unlike fluids, solids support both volumetric and shear deformation. Accordingly, elastic disturbances can propagate in longitudinal and transverse modes, associated with normal and tangential polarizations, respectively.

 For a longitudinal disturbance propagating along the \(x\)-direction, consider the strain tensor
\begin{equation}
\boldsymbol{\varepsilon}
 =
 \begin{pmatrix}
 e&0&0\\
 0&0&0\\
 0&0&0
 \end{pmatrix},
\end{equation}
where \(e=e(x)\). For constant Lamé parameters, the elastic contribution on the right-hand side of Eq.~\eqref{eq:31-000} becomes
\begin{equation}
 -(\lambda+2\mu)e\frac{\mathrm de}{\mathrm dx}.
\end{equation}
Using the one-dimensional relation between the velocity and density variations, the longitudinal density-evolution equation can therefore be written as
\begin{equation}
 v_p^2\,\mathrm d\rho
 =
 -(\lambda+2\mu)e\,\mathrm de
 =
 -(\lambda+2\mu)
 \boldsymbol{\varepsilon}:\mathrm d\boldsymbol{\varepsilon}.
\end{equation}

We introduce the following  compression relation for the normal deformation:
\begin{equation}
 \boldsymbol{\varepsilon}:
 \frac{\mathrm d\boldsymbol{\varepsilon}}{\mathrm d\rho}
 =-\frac{1}{\rho}.
\end{equation}
It then follows that
\begin{equation}
 v_p^2=\frac{\lambda+2\mu}{\rho}.
\end{equation}
Recalling the relation between the Lamé parameters and the bulk and shear moduli~\cite{Gourgiotis2016},
\begin{equation}
K=\lambda+\frac{2}{3}\mu,
\end{equation}
the longitudinal sound speed can be expressed as
\begin{equation}
 \boxed{
 v_p=\sqrt{\frac{K+\frac{4}{3}\mu}{\rho}}.
 }
\end{equation}
This is the classical longitudinal-wave speed for a homogeneous, isotropic, linear-elastic solid.

For transverse elastic disturbances, choose Cartesian coordinates such that the propagation direction is along the \(x\)-axis and the polarization direction is along the \(y\)-axis. We 
consider a shear-strain field of the form
\begin{equation}
\boldsymbol{\varepsilon}
=
\begin{pmatrix}
0&e/2&0\\
e/2&0&0\\
0&0&0
\end{pmatrix},
\qquad e=e(x).
\end{equation}
The trace of the strain tensor vanishes,
\begin{equation}
\operatorname{tr}\boldsymbol{\varepsilon}=0,
\end{equation}
indicating that the deformation is purely shear and involves no volumetric strain. The square of the strain tensor is
\begin{equation}
\boldsymbol{\varepsilon}^2
=
\begin{pmatrix}
e^2/4&0&0\\
0&e^2/4&0\\
0&0&0
\end{pmatrix},
\end{equation}
while its divergence is given by
\begin{equation}
\nabla\cdot\boldsymbol{\varepsilon}
=
\begin{pmatrix}
0\\
\frac12\frac{\mathrm de}{\mathrm dx}\\
0
\end{pmatrix}.
\end{equation}

Substituting these expressions into the density-evolution equation, the tangential component is obtained as
\begin{equation}
v_t\frac{\mathrm D\rho}{\mathrm Dt}
=
\frac{\mu e}{2}\frac{\mathrm de}{\mathrm dx}.
\end{equation}
For steady motion with a purely tangential velocity, the material density derivative is
\begin{equation}
\frac{\mathrm D\rho}{\mathrm Dt}
=
v_t\frac{\mathrm d\rho}{\mathrm dy},
\end{equation}
provided that the density varies along the tangential direction. Combining these relations yields
\begin{equation}
\boxed{
v_t^2\frac{\mathrm d\rho}{\mathrm dy}
=
\frac{\mu e}{2}\frac{\mathrm de}{\mathrm dx}.
}
\end{equation}

Defining the compression in the tangential direction
\begin{equation}
\frac{e}{2}
\frac{\mathrm de}{\mathrm d\rho}
=
\frac{1}{\rho}.
\end{equation}
we obtain the transverse-wave speed  as 
\begin{equation}
\boxed{
v_t=\sqrt{\frac{\mu}{\rho}}
=\sqrt{\frac{G}{\rho}},
}
\end{equation}
where \(G=\mu\) denotes the shear modulus. This result identifies the transverse-wave speed with the propagation of a purely shear disturbance, for which the restoring response is governed by the shear modulus.

\section{Galilei invariance and objectivity}

The objectivity of the system energy is a fundamental principle underlying the formulation of the present viscoelastic theory. In this work, the observer independence of the system energy is ensured by introducing a Lagrange multiplier  $P$ associated with the mass $m$. Accordingly, the energy under a Galilean transformation is expressed as
\begin{equation}
 E=mU^2+P,
\end{equation}
where \(U\) denotes the Galilean-transformation velocity, and \(P\) is the associated Lagrange multiplier. As demonstrated in Ref.~\cite{wang2026density}, the transformation energy is independent of the observer provided that \(\mathrm{D}U/\mathrm{D}t=0\).

We next clarify the role of the momentum evolution equation and its relation to objectivity. In the present framework, the system dynamics are governed by the velocity-diffusion equation, rather than directly by the momentum evolution equation. The latter primarily serves to determine the pressure-like Lagrange multiplier, rather than to independently determine the velocity field. This interpretation is reflected in the equivalence between the momentum evolution equation and the kinetic Gibbs--Duhem relation: $
 \mathrm{d}p+\mathbf{v}\cdot\mathrm{d}(\rho\mathbf{v})=0$.
Since the pressure is  intrinsically proportional to \(\mathbf{v}^2\)~\cite{wang2026density}, its differential is  proportional to  $\mathbf{v}\cdot\mathrm{d}\mathbf{v}$
and cancels the contribution of the observer-dependent velocity in the second term. Consequently, the kinetic Gibbs--Duhem relation, and hence the corresponding momentum evolution equation, is invariant under Galilean transformations.

The present viscoelastic theory does not introduce a Maxwell-type convective derivative~\cite{Oldroyd1950,Oldroyd1958} or an objective stress-rate formulation solely to enforce objectivity in the temporal evolution of the constitutive variables. Instead, the evolution equations are formulated to respect the underlying Galilean transformation properties of the system. In particular, both the velocity and strain evolution equations are Galilean-covariant, such that their physical content is preserved under a change of inertial reference frame.

\section{Conclusion}

In this work, we have developed an energy-based theoretical framework for describing coupled viscous and elastic dynamics in continuum materials. By incorporating kinetic energy and elastic deformation into a unified free-energy formulation, we derived coupled evolution equations for the velocity, strain, and momentum fields. The formulation separates dissipative and nondissipative contributions, allowing viscous effects and elastic responses to be described within a common continuum framework.

The dissipative dynamics give rise to a velocity-diffusion equation governed by the viscosity and density, together with a phenomenological strain-diffusion equation describing the evolution of elastic deformation. The nondissipative contribution determines the momentum evolution through the volumetric and shear components of the elastic response. This formulation provides a basis for investigating the interplay between viscous dissipation, elastic deformation, and momentum transport.

Under homogeneous, isotropic, linear-elastic, and isentropic assumptions, the framework also recovers the classical sound speeds in both liquids and solid materials. For a liquid, the longitudinal sound speed is given by
\[
v_l=\sqrt{\frac{K}{\rho}},
\]
whereas, for an isotropic linear-elastic solid, the longitudinal and transverse wave speeds are
\[
v_p=\sqrt{\frac{K+\frac{4}{3}G}{\rho}},
\qquad
v_s=\sqrt{\frac{G}{\rho}},
\]
respectively. Here, \(K\), \(G\), and \(\rho\) denote the bulk modulus, shear modulus, and mass density, respectively. These results demonstrate that the proposed formulation recovers the classical sound speed in liquids as well as the longitudinal and transverse elastic wave speeds in solids, thereby connecting fluid and solid dynamics within a unified framework.

In conclusion, the present work establishes an energy-based approach to coupled viscous–elastic continuum dynamics, combining velocity diffusion, strain evolution, elastic momentum transport, and wave propagation. The recovery of classical sound speeds in both liquids and solids illustrates the applicability of the framework across different material phases. Future work will focus on establishing the thermodynamic consistency of the coupled evolution equations, clarifying their constitutive assumptions, and investigating their predictions for dynamical processes in viscous–elastic materials.

\section*{Statement}

During the preparation of this work, the author used AI-assisted tools to improve grammar and language clarity, particularly to address challenges associated with writing in a non-native language. Following the use of these tools, the author carefully reviewed and edited the manuscript as necessary and takes full responsibility for the content of the published article.

\section*{Funding}
This research did not receive any specific grant from funding agencies in the public, commercial, or not-for-profit sectors.

\section*{Acknowledgments} 

I am grateful to R. Yin for many valuable discussions.

\end{document}